\documentclass[12pt]{article}

\usepackage{graphicx}
\usepackage{hyperref}
\usepackage{slashed}
\usepackage{amsmath}
\usepackage{amssymb}
\usepackage{epsfig}
\usepackage{afterpage}

\newcommand{\bcln}{b \to c l^- \bar{\nu}_{l}}
\def \lb{\Lambda_b}
\def \lc{\Lambda_c}
\def \lbt{\Lambda_b \to \Lambda_c \tau \bar{\nu}_{\tau}}
\def \lbl{\Lambda_b \to \Lambda_c \ell \bar{\nu}_{\ell}}

\def\beq{\begin{equation}}
\def\eeq{\end{equation}}
\def\bea{\begin{eqnarray}}
\def\eea{\end{eqnarray}}
\def\ber{\begin{eqnarray*}}
\def\eer{\end{eqnarray*}}
\def\bwt{\begin{widetext}}
\def\ewt{\end{widetext}}
\def\nn{\nonumber}
\def\roughly#1{\mathrel{\raise.3ex\hbox
{$#1$\kern-.75em\lower1ex\hbox{$\sim$}}}}

\def\order{\lower 1.8ex \hbox{\LARGE\~{}}}

\def\BDtaunu{\bar{B} \to D^+ \tau^{-} {\bar\nu}_\tau}
\def\BDlnu{\bar{B} \to D^+ \ell^{-} {\bar\nu}_\ell}
\def\BDstartaunu{\bar{B} \to D^{*+} \tau^{-} {\bar\nu}_\tau}
\def\BDstarlnu{\bar{B} \to D^{*+} \ell^{-} {\bar\nu}_\ell}

\newcommand{\bctaunutau}{b \to c \tau^- {\bar\nu}_\tau}

\def\bra#1{\left\langle #1\right|}
\def\ket#1{\left| #1\right\rangle}

\def \({\left(}
\def \){\right)}
\def \[{\left[}
\def \]{\right]}
\def \l|{\left|}
\def \r|{\right|}

\def \nn{\nonumber}
\def \nl{\nn \\}

\begin{document}

\begin{flushright}
UMISS-HEP-2015-01 \\
\end{flushright}

\begin{center}
\bigskip
{\Large \bf \boldmath $\lbt$ Decay in the Standard Model and with New Physics} \\
\bigskip
\bigskip
{\large
Shanmuka Shivashankara $^{}$\footnote{sshivash@go.olemiss.edu},
Wanwei Wu $^{}$\footnote{wwu1@go.olemiss.edu} and
Alakabha Datta $^{}$\footnote{datta@phy.olemiss.edu} \\
}
\end{center}
\begin{flushleft}

~~~~~~~~~~~ {\it Department of Physics and
  Astronomy, 108 Lewis Hall, }\\
~~~~~~~~~~~~~~~{\it University of Mississippi, Oxford, MS
  38677-1848, USA }
\end{flushleft}

\begin{center}
\bigskip (\today)
\vskip0.5cm {\Large Abstract\\} \vskip3truemm
\parbox[t]{\textwidth}{Recently hints of
 lepton flavor non-universality emerged when the BaBar Collaboration 
 observed deviations from the standard model predictions in $R(D^{(*)}) \equiv
  {\cal B}({\bar B} \to D^{(*)+} \tau^- {\bar\nu}_\tau) / {\cal
    B}({\bar B} \to D^{(*)+} \ell^- {\bar\nu}_\ell)$ ($\ell =
  e,\mu$). Another test of this  non-universality can be in the semileptonic
$\lbt$ decay. In this work  we present predictions for this decay in the standard model
and in the presence of
  new-physics operators with different Lorentz structures. We present the most general four-fold angular distribution for this decay including new physics. For phenomenology, we focus on
predictions for the decay rate and the differential distribution in the momentum transfer
squared $q^2$. In particular, we calculate $R_{\lb}  =  \frac{BR[ \lbt]}{BR[\lbl]}$ where
 $\ell$ represents $\mu$ or $e$, and find the standard model prediction to be around $0.3$
while the new physics operators can increase or slightly decrease this value.
}
\end{center}

\thispagestyle{empty}
\newpage
\setcounter{page}{1}
\baselineskip=14pt

\section{Introduction}
A major part of particle physics research is focused on finding physics beyond the standard model (SM). In the flavor sector a key property of the SM gauge interactions is that they are lepton flavor universal. Evidence for violation of this property would be a clear sign of new physics (NP) beyond the SM. 
In the search for NP, the second and third generation quarks and leptons are quite special because they are comparatively heavier and are 
expected to be relatively more sensitive to NP. 
As an example, in certain versions of the two Higgs doublet models (2HDM) the couplings of the new Higgs bosons are proportional to the masses and so NP effects are more pronounced for the heavier generations. Moreover, the constraints on new physics, especially involving the third generation leptons and quarks, are somewhat weaker allowing for larger new physics effects.

Recently, the BaBar
Collaboration with their full data sample has reported the
following measurements \cite{RDexpt1,RDexpt2}:
\begin{eqnarray}
\label{babarnew}
R(D) &\equiv& \frac{{\cal B}(\BDtaunu)}
{{\cal B}(\BDlnu)}=0.440 \pm 0.058 \pm 0.042 ~, \nn \\
R(D^*) &\equiv& \frac{{\cal B}(\BDstartaunu)}
{{\cal B}(\BDstarlnu)} = 0.332 \pm 0.024 \pm 0.018 ~,
\label{RDexpt}
\end{eqnarray}
where $\ell = e,\mu$. The SM predictions are $R(D) = 0.297 \pm 0.017$
and $R(D^*) = 0.252 \pm 0.003$ \cite{RDexpt1,RDtheory}, which deviate
from the BaBar measurements by 2$\sigma$ and 2.7$\sigma$,
respectively. (The BaBar Collaboration itself reported a 3.4$\sigma$
deviation from SM when the two measurements of Eq.~(\ref{babarnew})
are taken together.) This  measurement of lepton flavor non-universality,
referred to as the $R(D^{(*)})$
puzzles, may be providing a hint of the new physics (NP) believed to
exist beyond the SM.
There have been numerous analyses examining NP explanations of the
$R(D^{(*)})$ measurements \cite{RDtheory,RDNP,dattaD,Bhattacharya:2014wla}.

The underlying quark level transition $\bctaunutau$ can be probed in both
$B$ and $\Lambda_b$ decays. Note that in the presence of lepton non-universality
the flavor of the neutrino does not have to match the flavor of the charged lepton \cite{Bhattacharya:2014wla}. Moreover the NP can affect all the lepton flavors. The main assumption
here is that the NP effect is largest for the $\tau$ sector and for simplicity we neglect
the smaller NP effects in the $\mu$ and $e$ leptons. 
The $\Lambda_b$ being a spin 1/2 baryon has a  complex angular distribution for its decay products. As in $B$ decays, we can construct several observables from
the angular distribution of the $\Lambda_b$ decay which can be used to find evidence of NP and to probe the structure of NP.

The decay $\lbt$ has not been measured experimentally though it might be possible to observe this decay at the LHCb. The full angular distribution of this decay is experimentally challenging and so
in this paper, for the sake of phenomenology, we will focus on the rate as well as the $q^2$ differential distribution for this decay. Using constraints on the new physics couplings
obtained by using Eq.~(\ref{babarnew}) we will make predictions for the effects of these couplings in
$\lbt$ decay. 
Recently, in Ref.~\cite{Gutsche:2015mxa} this decay was discussed in the standard model and with new physics in Ref.~\cite{Woloshyn:2014hka}.

The main uncertainty in the $\lbt$ decays are the hadronic form factors for the 
$\Lambda_b \to \Lambda_c$ transition. These form factors can also be studied systematically in
a heavy $m_b$ and $m_c$ expansion \cite{Datta:1994ij}. However, unlike the $B$ system
the heavy baryon form factors have not been extensively studied. We will therefore construct ratios where the form
factor uncertainties will mostly cancel leaving behind a smaller uncertainty for the theoretical predictions. We will then investigate if the NP effects are large enough to produce observable deviations from the SM predictions.

The paper is organized in the following manner. In sec.2  we introduce the effective Lagrangian to parametrize the NP operators,  describe the formalism of the decay process and introduce the relevant observables. In sec.3 we present our results and in sec.4 we present our conclusions.

\section{Formalism}
In the presence of NP, the effective Hamiltonian for the quark-level transition $\bcln$  can be written in the form \cite{ccLag}
\bea
\label{eq1:Lag}
{\cal{H}}_{eff} &=&  \frac{G_F V_{cb}}{\sqrt{2}}\Big\{
\Big[\bar{c} \gamma_\mu (1-\gamma_5) b  + g_L \bar{c} \gamma_\mu (1-\gamma_5)  b + g_R \bar{c} \gamma_\mu (1+\gamma_5) b\Big] \bar{l} \gamma^\mu(1-\gamma_5) \nu_l \nl && +  \Big[g_S\bar{c}  b   + g_P \bar{c} \gamma_5 b\Big] \bar{l} (1-\gamma_5)\nu_l + h.c \Big\}. \
\eea 
where  $G_F = 1.1663787 \times 10^{-5}\ GeV^{-2}$ is the Fermi coupling constant, $V_{cb}$ is the Cabibbo-Koboyashi-Maskawa (CKM) matrix element and we use $\sigma_{\mu \nu} = i[\gamma_\mu, \gamma_\nu]/2$. 
We have assumed the neutrinos to be always left chiral and to introduce non-universality the NP couplings are in general different for different lepton flavors. We assume the NP effect is mainly through the $\tau$ lepton and do not consider tensor operators in our analysis.
Further, we do not assume any relation between $b \to u l^- \bar{\nu}_l$ and $\bcln$ transitions and hence  do not include constraints from
$B \to \tau \nu_{\tau}$. 
The SM  effective Hamiltonian corresponds to $g_L = g_R = g_S = g_P = 0$.  

In Ref.\cite{dattaD} we had parametrized the NP in terms of the couplings $g_S$, $g_P$, $g_V=g_R+g_L$ and $g_A=g_R-g_L$ while in this work we have traded $g_V$ and $g_A$ for $g_{L,R}$ to align our analysis closer to realistic models \cite{Bhattacharya:2014wla}.  The couplings $g_{L,R,P}$ contribute to $R(D^*)$ while $g_{L,R,S}$ contribute to $R(D)$. We will consider one NP coupling at a time and
provide constraints on these couplings from $R(D^{(*)})$. 

\subsection{Decay Process}
The process under consideration is
$$\Lambda_{b}(p_{\lb})\rightarrow\tau^{-}(p_{1})+\bar{\nu_{\tau}}(p_{2})+\Lambda_{c}(p_{\lc})$$
In the SM the amplitude for this process is
\bea
M_{SM}&=&\frac{G_{F}V_{cb}}{\sqrt{2}}L^{\mu}H_{\mu},\
\eea
where the leptonic and hadronic currents are,
\bea
L^{\mu}&=&\bar{u}_{\tau}(p_{1})\gamma^{\mu}(1-\gamma_{5})v_{\nu_{\tau}}(p_{2}), \nonumber\\
H_{\mu} & = &\bra{\lc}\bar{c}\gamma_{\mu}(1-\gamma_5)b\ket{\lb}. \
\eea
The hadronic current is expressed in terms of six form factors,
\bea
\bra{\lc}\bar{c}\gamma_{\mu}b\ket{\lb} & = & \bar{u}_{\Lambda_{c}}(f_{1}\gamma_{\mu}+i f_{2}\sigma_{\mu\nu}q^{\nu}+f_{3}q_{\mu})u_{\Lambda_{b}}, \nonumber\\
\bra{\lc}\bar{c}\gamma_{\mu}\gamma_{5}b|\ket{\lb} &= & \bar{u}_{\Lambda_{c}}(g_{1}\gamma_{\mu}\gamma_{5}+i g_{2}\sigma_{\mu\nu}q^{\nu}\gamma_{5}+g_{3}q_{\mu}\gamma_{5})u_{\Lambda_{b}}.\
\eea
Here $q= p_{\lb}- p_{\lc}$ is the momentum transfer and the form factors are functions of $q^2$.
When considering NP operators we will use the following relations obtained by using the equations of motion.
\bea
\bra{\lc}\bar{c}b\ket{\lb} &= & \bar{u}_{\Lambda_{c}}(f_{1}\frac{\slashed{q}}{m_b-m_{c}}+f_{3}\frac{q^2}{m_b-m_{c}})u_{\Lambda_{b}}, \nonumber\\
\bra{\lc}\bar{c}\gamma_{5}b\ket{\lb} &= &\bar{u}_{\Lambda_{c}}(-g_{1}\frac{\slashed{q}\gamma_{5}}{m_b+m_{c}}-g_{3}\frac{q^2\gamma_{5}}{m_b+m_{c}})u_{\Lambda_{b}}.\
\eea

We will define the following observable,
\bea
\label{ratio1}
R_{\lb} & = & \frac{BR[ \lbt]}{BR[\lbl]}. \
\eea
Here $\ell$ represents $\mu$ or $e$. We will also define the ratio of differential distributions,
\bea
\label{ratio2}
B_{\lb}(q^2) & = & \frac{\frac{d\Gamma[ \lbt]}{d q^2}}{\frac{ d \Gamma[\lbl]}{d q^2}}. \
\eea

Our results will show that these observables are not very sensitive to variations in the hadronic form
factors.

\subsection{ Helicity Amplitudes and the Full Angular Distribution}
The decay $\lbt$ proceeds via $\Lambda_b\rightarrow\Lambda_c
W^*$({off-shell} W) followed by $W^*\rightarrow \tau \bar{\nu}_\tau$. The full decay process is $\lb \to \lc ( \to \Lambda_s \pi) W^* ( \to \tau \bar{\nu}_{\tau}) $
Following
\cite{Korner:1991ph} one can analyze the decay in terms of  helicity amplitudes
which are given  by
\begin{equation}
\label{heldef}
H_{\lambda_2\lambda_W} =M_\mu(\lambda_2)\epsilon^{*\mu}(\lambda_W),
\end{equation}
where $\lambda_2,\lambda_{W}$ are the polarizations of the daughter baryon and
the W-boson respectively and $M_\mu$ is the hadronic current for $\lb \to \lc$ transition. The helicity amplitudes can be expressed in terms of form factors and the NP couplings.  
\begin{eqnarray}
\label{helicity1}
H_{\lambda_{\Lambda_{c}},\lambda_{w}}&=&H_{\lambda_{\Lambda_{c}},\lambda_{w}}^V -H_{\lambda_{\Lambda_{c}},\lambda_{w}}^A,\nonumber\\
H_{\frac120}^{V}&=&(1 + g_L + g_R) \frac{\sqrt{Q_-}}{\sqrt{q^2}}
  \bigg((M_1 + M_2)f_1 - q^2f_2\bigg),\nonumber\\
  H_{\frac120}^{A}&=&(1 + g_L - g_R)\frac{\sqrt{Q_+}}{\sqrt{q^2}}
  \bigg((M_1 - M_2)g_1 + q^2g_2\bigg),\nonumber\\
H_{\frac121}^{V}&=&(1 + g_L + g_R)\sqrt{2Q_-}
  \bigg(f_1 - (M_1 + M_2)f_2\bigg),\nonumber\\
H_{\frac121}^{A}&=&(1 + g_L - g_R)\sqrt{2Q_+}
  \bigg(g_1 + (M_1 - M_2)g_2\bigg),\nonumber\\ 
H_{\frac12t}^{V}&=&(1 + g_L + g_R)\frac{\sqrt{Q_+}}{\sqrt{q^2}}
  \bigg((M_1 - M_2)f_1+ q^2f_3\bigg),\nonumber\\
   H_{\frac12t}^{A}&=&(1 + g_L - g_R)\frac{\sqrt{Q_-}}{\sqrt{q^2}}
  \bigg((M_1 + M_2)g_1 - q^2g_3\bigg),
\end{eqnarray}
where $Q_{\pm}= (M_1 \pm M_2)^2-q^2$.

We also have,
\begin{eqnarray}
H_{\lambda_{\Lambda_{c}},\lambda_{w}}^V&=&H_{-\lambda_{\Lambda_{c}},-\lambda_{w}}^V,\nonumber\\
H_{\lambda_{\Lambda_{c}},\lambda_{w}}^A&=&-H_{-\lambda_{\Lambda_{c}},-\lambda_{w}}^A.
\end{eqnarray}

The scalar and pseudo-scalar helicities associated with the new physics scalar and pseudo-scalar interactions are
\bea
\label{SP}
{H^{SP}}_{1/2,0} \  & = & \ {H^{P}}_{1/2,0} + {H^{S}}_{1/2,0},\nonumber\\
{H^{S}}_{1/2,0}  & = & g_S \frac{\sqrt{Q_+}}{m_b-m_c}\Big( (M_1-M_2)f_1 +q^2 f_3\Big),\nonumber\\
{H^{P}}_{1/2,0} &= & -g_P \frac{\sqrt{Q_-}}{m_b+m_c}\Big( (M_1+M_2)g_1 - q^2 g_3\Big).\
\eea

The parity related amplitudes are,
\bea
{H^{S}}_{\lambda_{\Lambda_{c}},\lambda_{NP}} & = & {H^{S}}_{-\lambda_{\Lambda_{c}},-\lambda_{NP}},\nonumber\\
{H^{P}}_{\lambda_{\Lambda_{c}},\lambda_{NP}} & = & -{H^{P}}_{-\lambda_{\Lambda_{c}},-\lambda_{NP}}.\
\eea

With the W boson momentum defining the positive z-axis for the decay process ($\Lambda_{b}\rightarrow \Lambda_{c} \tau^{-} \nu_{\tau}$), the twofold angular distribution can be written as
\bea
\label{dqc}
\frac{d\Gamma(\Lambda_{b}\rightarrow \Lambda_{c} \tau^{-} \nu_{\tau})}{dq^2 d(\cos{\theta_l})} &=& \frac{{G_F}^2 {|V_{cb}}|^2 q^2 |  {\bf p}_{\Lambda_{c}}|}{512 \pi^3 {M_1}^2} \left(1-\frac{{m_l}^2}{q^2}\right)^2 \Big[ A_{1}^{SM} +  \frac{{m_l}^2}{q^2} A_{2}^{SM} + 2 A_{3}^{NP}\nl && + \frac{4{m_l}}{\sqrt{q^2}} A_{4}^{Int} \Big]
\eea
\\
where,
\begin{eqnarray}
A_{1}^{SM}&=&2 \sin^2{\theta_l} (| H_{1/2,0} |^2 + | H_{-1/2,0} |^2  ) + (1-\cos{\theta_l})^2 | H_{1/2,1} |^2 \nl && + (1+\cos{\theta_l})^2 | H_{-1/2,-1} |^2,\nonumber\\
A_{2}^{SM}&=&2 \cos^2{\theta_l} (| H_{1/2,0} |^2  + | H_{-1/2,0} |^2 ) + \sin^2{\theta_l} (| H_{1/2,1} |^2 + | H_{-1/2,-1} |^2 ) \nl && + 2 (| H_{1/2,t} |^2 + | H_{-1/2,t} |^2 )  - 4 \cos{\theta_l} Re[(H_{1/2,t} \  (H_{1/2,0})^* + H_{-1/2,t} \ (H_{-1/2,0})^*)],\nonumber\\
A_{3}^{NP}&=&| {H^{SP}}_{1/2,0} |^2 + | {H^{SP}}_{-1/2,0} |^2 ,\nonumber\\
A_{4}^{Int}&=&-\cos{\theta_l} Re[(H_{1/2,0} \ ({H^{SP}}_{1/2,0})^* + H_{-1/2,0} \ ({H^{SP}}_{-1/2,0})^*)] \nl && + Re[(H_{1/2,t} \ ({H^{SP}}_{1/2,0})^* + H_{-1/2,t} \ ({H^{SP}}_{-1/2,0})^*)].
\end{eqnarray}
\noindent
$A{_1}^{SM}$, $A{_2}^{SM}$, $A{_3}^{NP}$, and $A{_4}^{Int}$ are the standard model non-spin-flip, standard model spin-flip, new physics, and interference terms, respectively apart from $g_L$ and $g_R$. Note $A{_1}^{SM}$, $A{_2}^{SM}$ have the same structure as the SM contributions but the helicity amplitudes in these quantities include the new physics contributions from $g_{L,R}$.
$\theta_l$ is the angle of the lepton in the W rest frame with respect to the W momentum.

 After integrating out $cos{\theta_l}$,

\bea
\label{dq1}
\frac{d\Gamma(\Lambda_{b}\rightarrow \Lambda_{c} \tau^{-} \nu_{\tau})}{dq^2} &=&
\frac{{G_F}^2 {|V_{cb}}|^2 q^2 |  {\bf p}_{\Lambda_{c}}|}{192\pi^3 {M_1}^2} \left(1-\frac{{m_l}^2}{q^2}\right)^2 \Big[ B_{1}^{SM} + \frac{{m_l}^2}{2q^2} B_{2}^{SM} + \frac{3}{2} B_{3}^{NP}\nl && + \frac{3{m_l}}{\sqrt{q^2}} B_{4}^{Int}\Big]
\eea
\\
where,
\begin{eqnarray}
B_{1}^{SM}&=&| H_{1/2,0} |^2  + | H_{-1/2,0} |^2  +  | H_{1/2,1} |^2 + | H_{-1/2,-1} |^2 ,\nonumber\\
B_{2}^{SM}&=&| H_{1/2,0} |^2 + | H_{-1/2,0} |^2  + | H_{1/2,1} |^2 + | H_{-1/2,-1} |^2\nl && + 3 (| H_{1/2,t} |^2 + | H_{-1/2,t} |^2 ),\nonumber\\
B_{3}^{NP}&=& | H^{SP}_{1/2,0} |^2 + | H^{SP}_{-1/2,0} |^2 ,\nonumber\\
B_{4}^{Int}&=&Re[(H_{1/2,t} \ ({H^{SP}}_{1/2,0})^*  + H_{-1/2,t} \ ({H^{SP}}_{-1/2,0})^*)].\\\nonumber
\end{eqnarray}
\noindent
$B{_1}^{SM}$, $B{_2}^{SM}$, $B{_3}^{NP}$, and $B{_4}^{Int}$ are the standard model non-spin-flip, standard model spin-flip, new physics, and interference terms, respectively apart from $g_L$ and $g_R$.  Again, $B{_1}^{SM}$, $B{_2}^{SM}$ have the same structure as the SM contributions but the helicity amplitudes in these quantities include the new physics contributions from $g_{L,R}$. The $g_{S,P}$ operators generate new terms in the angular distribution.

The angular distribution for the four body decay process \Big($\Lambda_{b}\rightarrow (\Lambda_{s}, \pi^{+}) \Lambda_{c} \tau^{-} \nu_{\tau}$\Big) can be written as the following where $\alpha$ is the parity parameter for the process $\Lambda_{c}\rightarrow \Lambda_{s}\pi^{+}$.   $\theta_l$ is again the same leptonic angle.   $\theta_s$ is the angle of $\Lambda_s$ in the $\Lambda_c$ rest frame with respect to the $\Lambda_{c}$ momentum.  $\chi$ is the dihedral angle between the decay planes of ($\tau^-,\nu_{\tau}$) and ($\Lambda_s,\pi^+$) in the W and $\Lambda_c$ rest frame, respectively.

\bea
\label{dqcphi}
\frac{d\Gamma(\Lambda_{b}\rightarrow (\Lambda_{s}, \pi^{+}) \Lambda_{c} \tau^{-} \nu_{\tau})}{dq^2 d(\cos{\theta_l}) d\chi d(\cos{\theta_s})} &=& \frac{{G_F}^2 {|V_{cb}}|^2 q^2 | {\bf p}_{\Lambda_{c}}|}{2^{7} (2\pi)^4 {M_1}^2} \left(1-\frac{{m_l}^2}{q^2}\right)^2 \Big[C_{1}^{SM} + \frac{{m_l}^2}{q^2} C_{2}^{SM} + 2C_{3}^{NP}\nl && + \frac{4{m_l}}{\sqrt{q^2}} C_{4}^{Int}\Big]
\eea \\*
where,
\begin{eqnarray*}
C_{1}^{SM}&=&2 \sin^2{\theta_l} \Big((1+\alpha \cos{\theta_s}) | H_{1/2,0} |^2   + (1-\alpha \cos{\theta_s}) | H_{-1/2,0} |^2 \Big) \nl && + (1+\cos{\theta_l})^2 (1-\alpha \cos{\theta_s}) | H_{-1/2,-1} |^2  + (1-\cos{\theta_l})^2 (1+\alpha \cos{\theta_s}) | H_{1/2,1} |^2 \nl && - \frac{4\alpha}{\sqrt{2}} \sin{\theta_l} \sin{\theta_s} \cos{\chi}  \Big( (1+\cos{\theta_l})  Re[H_{1/2,0} \ (H_{-1/2,-1})^*] \nl && + (1-\cos{\theta_l}) Re[H_{-1/2,0} \ (H_{1/2,1})^*]  \Big) \nl &&  - \frac{4\alpha}{\sqrt{2}} \sin{\theta_l} \sin{\theta_s} \sin{\chi}  \Big( (1+\cos{\theta_l})  Im[H_{1/2,0} \ (H_{-1/2,-1})^*] \nl && - (1-\cos{\theta_l}) Im[H_{-1/2,0} \ (H_{1/2,1})^*]  \Big).\nonumber\\
\end{eqnarray*}
\begin{eqnarray*}
C_{2}^{SM}&=&2 \cos^2{\theta_l} \Big((1+\alpha \cos{\theta_s}) | H_{1/2,0} |^2  + (1-\alpha \cos{\theta_s}) | H_{-1/2,0} |^2 \Big) \nl && + \sin^2{\theta_l} \Big((1+\alpha \cos{\theta_s}) | H_{1/2,1} |^2 + (1-\alpha \cos{\theta_s}) | H_{-1/2,-1} |^2 \Big) \nl && + \frac{2\alpha}{\sqrt{2}} \sin{2\theta_l} \sin{\theta_s} \cos{\chi} \Big(Re[H_{1/2,0} \ (H_{-1/2,-1})^*]  - Re[H_{-1/2,0} \ (H_{1/2,1})^*]\Big) \nl && + \frac{2\alpha}{\sqrt{2}} \sin{2\theta_l} \sin{\theta_s} \sin{\chi} \Big(Im[H_{1/2,0} \ (H_{-1/2,-1})^*]  + Im[H_{-1/2,0} \ (H_{1/2,1})^*]\Big) \nl && - 4\cos{\theta_l} \Big((1+\alpha \cos{\theta_s}) Re[H_{1/2,t} \ (H_{1/2,0})^*]  + (1-\alpha \cos{\theta_s})  Re[H_{-1/2,t} \ (H_{-1/2,0})^*]\Big)\nl && - \frac{4\alpha}{\sqrt{2}} \sin{\theta_l} \sin{\theta_s} \cos{\chi} \Big(Re[H_{1/2,t} \ (H_{-1/2,-1})^*]  - Re[H_{-1/2,t} \ (H_{1/2,1})^*]\Big) \nl && - \frac{4\alpha}{\sqrt{2}} \sin{\theta_l} \sin{\theta_s} \sin{\chi} \Big(Im[H_{1/2,t} \ (H_{-1/2,-1})^*]  + Im[H_{-1/2,t} \ (H_{1/2,1})^*]\Big) \nl && + 2 \Big((1+\alpha \cos{\theta_s}) | H_{1/2,t} |^2 + (1-\alpha \cos{\theta_s}) | H_{-1/2,t} |^2 \Big).\nonumber\\
\end{eqnarray*}
\begin{eqnarray}
C_{3}^{NP}&=&(1+\alpha \cos{\theta_s}) | {H^{SP}}_{1/2,0} |^2 + (1-\alpha \cos{\theta_s}) | {H^{SP}}_{-1/2,0} |^2,\nonumber\\
C_{4}^{Int}&=&-\cos{\theta_l}\Big( (1+\alpha \cos{\theta_s}) Re[H_{1/2,0} \ ({H^{SP}}_{1/2,0})^*] \nl &&  + (1-\alpha \cos{\theta_s}) Re[H_{-1/2,0} \ ({H^{SP}}_{-1/2,0})^*]\Big) \nl && + (1+\alpha \cos{\theta_s})  Re[H_{1/2,t} \ ({H^{SP}}_{1/2,0})^*] \nl && + (1-\alpha \cos{\theta_s}) Re[H_{-1/2,t} \ ({H^{SP}}_{-1/2,0})^*].
\end{eqnarray}
$C{_1}^{SM}$, $C{_2}^{SM}$, $C{_3}^{NP}$, and $C{_4}^{Int}$ are the standard model non-spin-flip, standard model spin-flip, new physics, and interference terms, respectively apart from $g_L$ and $g_R$.
$C{_1}^{SM}$ and $C{_2}^{SM}$ have the same structure as the SM contributions but the helicity amplitudes in these quantities include the new physics contributions from $g_{L,R}$.  Several additional observables can be constructed from the angular distributions, such as polarization asymmetries and CP violating triple product asymmetries \cite{TP} which can be sensitive probes of new physics. Note that the standard model portion of the twofold and fourfold distributions above, Eq.~\ref{dqc} and Eq.~\ref{dqcphi}, are the same as in a recent paper\cite{Gutsche:2015mxa} apart from a minus sign in $C_2^{SM}$ above. $^{}$\footnote{In \cite{Gutsche:2015mxa} Eq. (51), the minus sign is required in front of sin2$\theta$ on the second line  in the spin-flip term as can be seen by the $d-matrix$ elements.} 

\section{Numerical Results}

\subsection{ New Physics Couplings}
\begin{figure}
\begin{center}
\includegraphics[width=6.5cm, height=5.5cm]{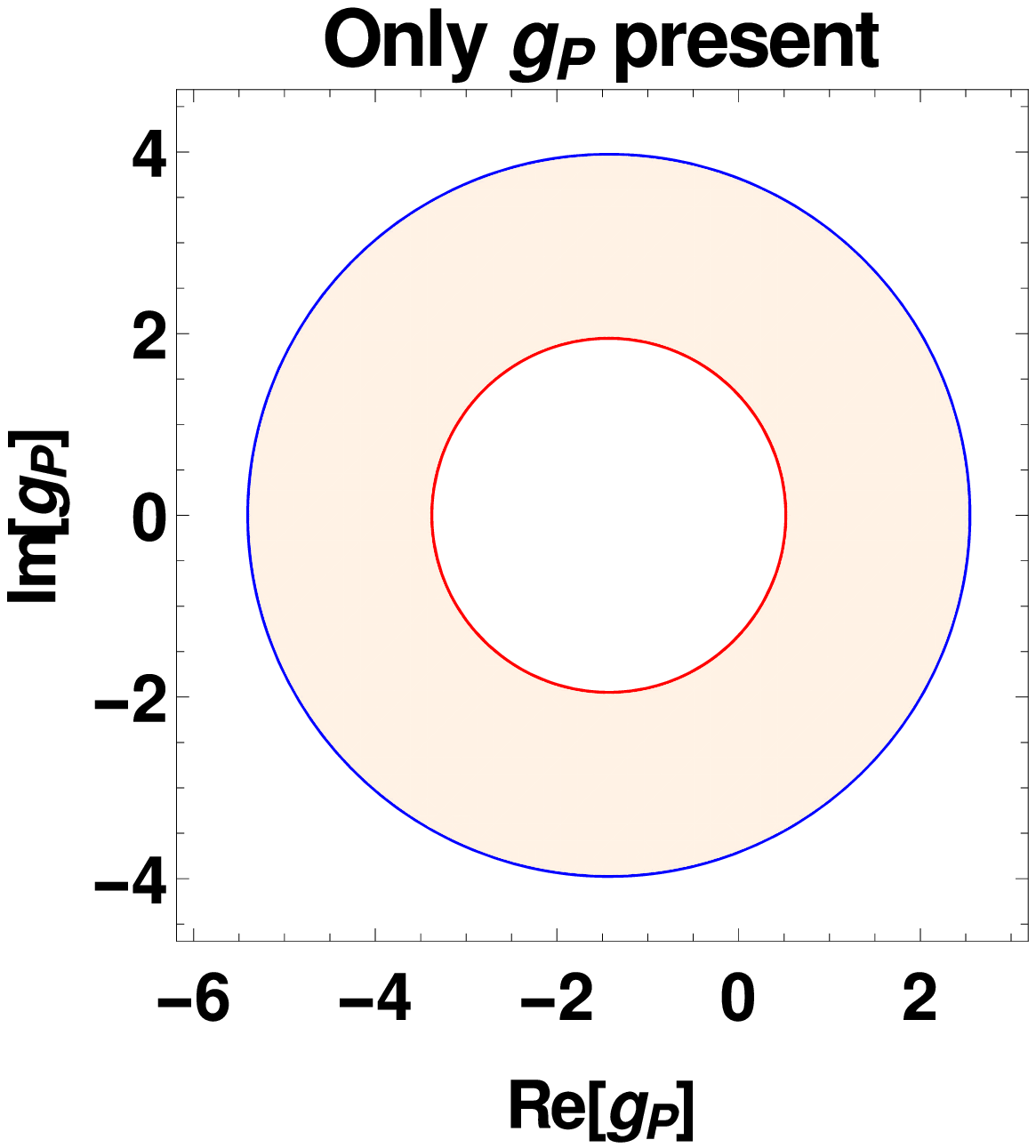}
\includegraphics[width=6.5cm, height=5.5cm]{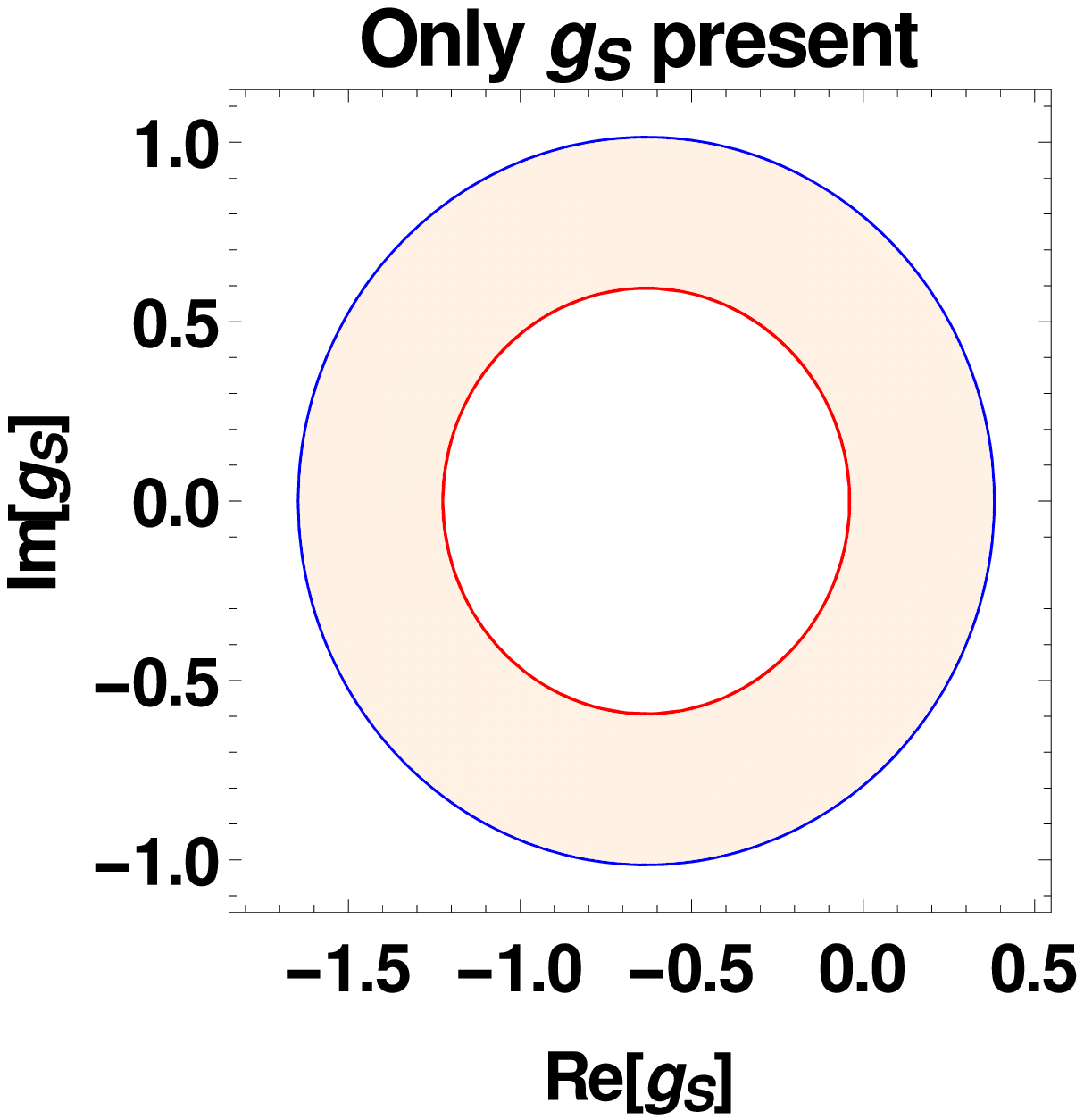}
\includegraphics[width=6.5cm, height=5.5cm]{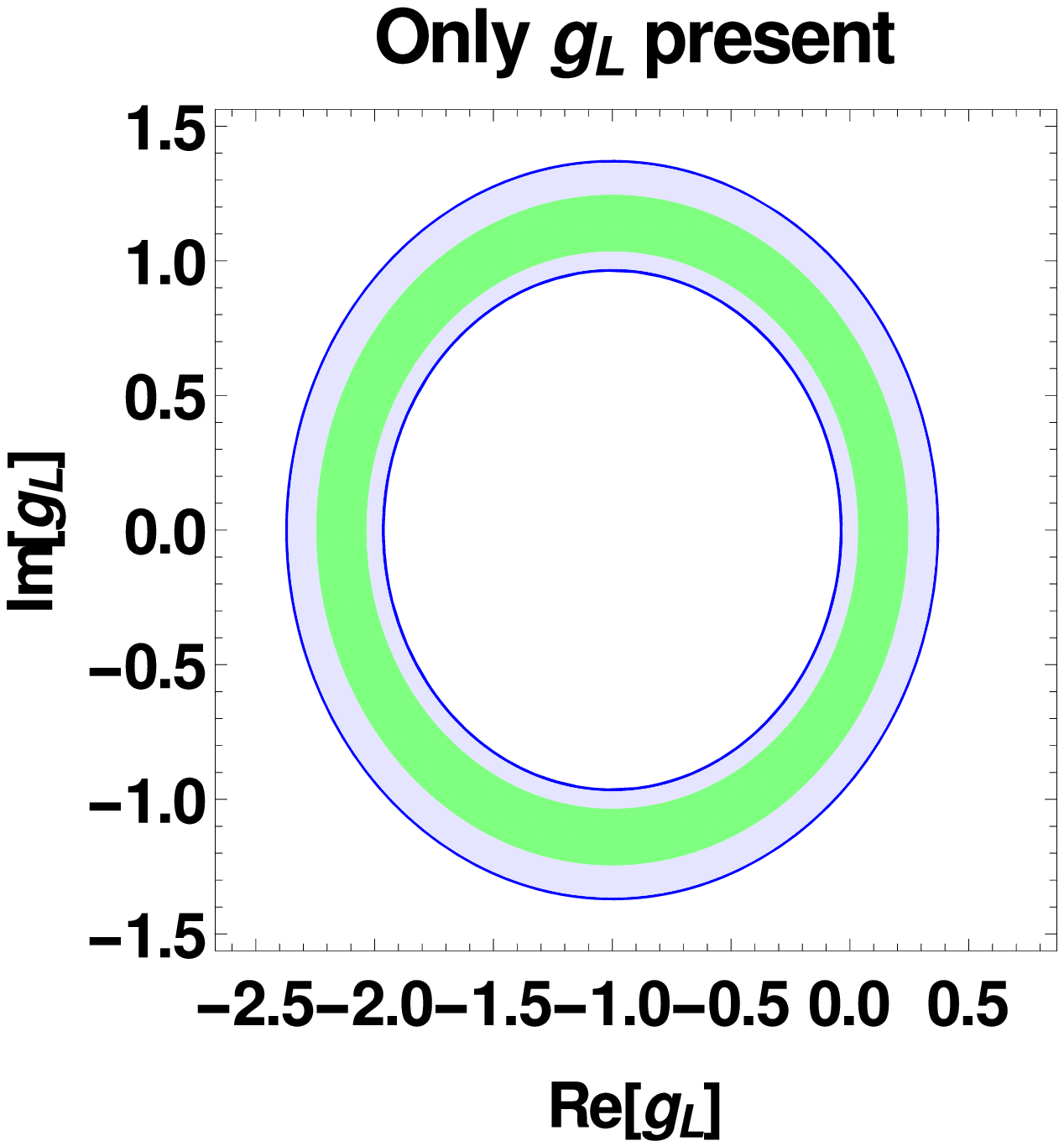}
\includegraphics[width=6.5cm, height=5.5cm]{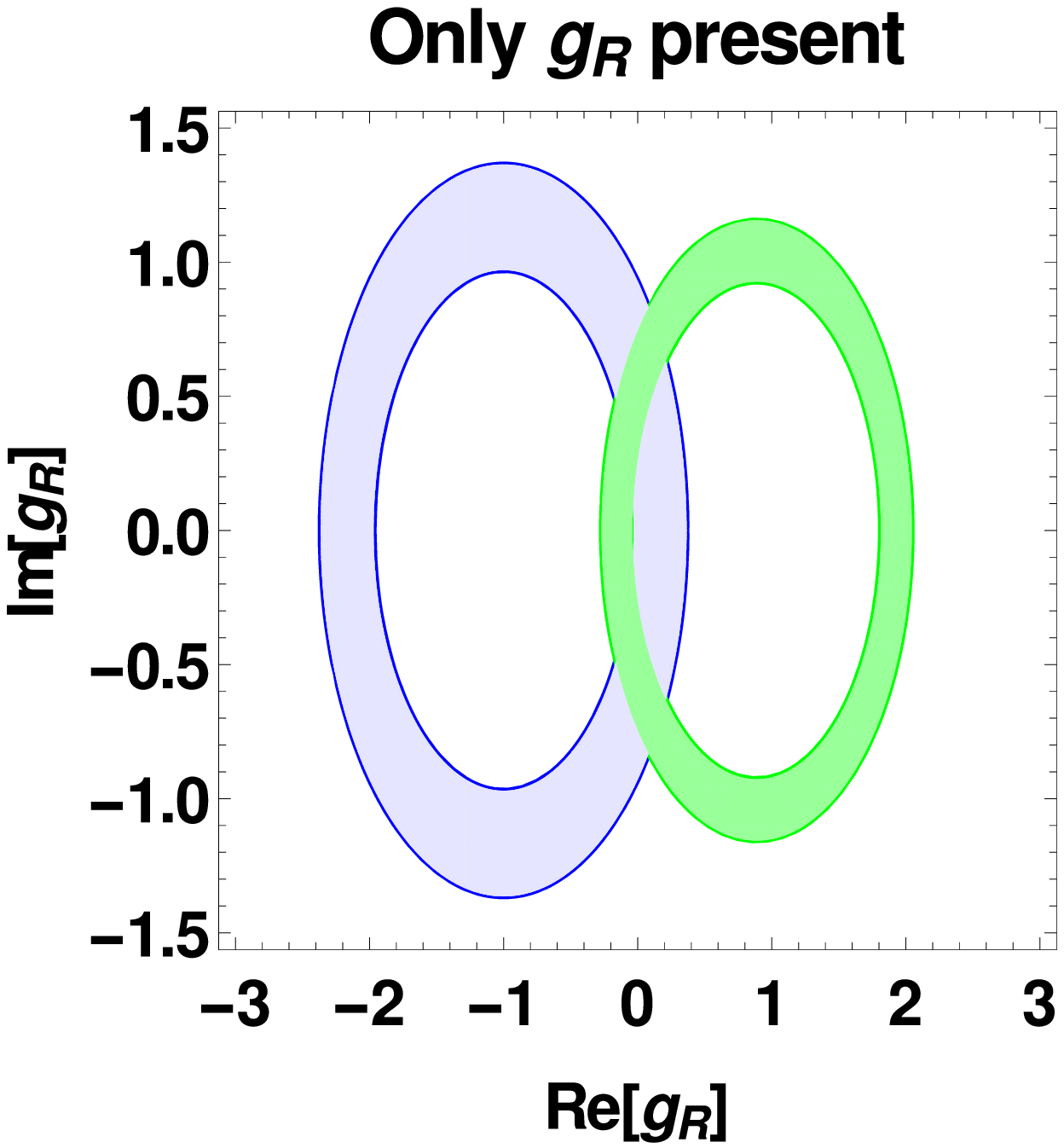}
\end{center}
\caption{The figures show the constraints on the NP couplings taken one at a time at the 95\% CL limit \cite{RDtheory,dattaD}. When the 
couplings contribute to both $R(D)$  and $R(D^*)$ the green contour indicates constraint from $R(D^*)$ and blue from
$R(D)$.}
\label{contour}
\end{figure}

We first present the constraints on the NP couplings from $R(D^{(*)})$. The couplings $g_S$ only contributes to $R(D)$, 
 $g_P$ only contributes to $R(D^*)$ while  $g_{L,R}$  contributes to both $R(D)$  and $R(D^*)$. The details of the calculations for Fig.~\ref{contour} can be found in
Ref.~\cite{RDtheory,dattaD}.

\subsection{Form Factors}
One of the main inputs in our calculations are the form factors. As first principle, lattice calculations of the form factors are not yet available.
The form factors we use here are from QCD sum rules, which is a well known approach to compute non-perturbative effects like form factors for systems with both light and heavy quarks\cite{FF1,FF2}.

In Ref.~\cite{FF2}, various parametrizations of the form factors are used. They are shown in Table. 1 ( $t=q^2$).

\begin{table}[tbh]
\center
\begin{tabular}{|c|c|c|c|}\hline
continuum \ model & $\kappa$ & $F_1^V(t)=f_1$ & $F_2^V(t)(GeV^{-1})=f_2$\\
\hline
rectangular & 1 & ${6.66/(20.27-t)}$ & ${-0.21/(15.15-t)}$ \\
rectangular & 2 & ${8.13/(22.50-t)}$ & ${-0.22/(13.63-t)}$\\
triangular & 3 & ${13.74/(26.68-t)}$ & ${-0.41/(18.65-t)}$\\
triangular & 4 & ${16.17/(29.12-t)}$ & ${-0.45/(19.04-t)}$\\
\hline
\end{tabular}
\caption{Various choices of Form Factors.}
\label{FF}
\end{table}

\vskip0.2cm

The form factors satisfy the heavy quark effective theory relations in the $m_b \to \infty$ limit.
\beq
f_{1}=g_{1} \quad f_{2}=g_{2} \quad f_{3}=g_{3}=0.\
\eeq

\subsection{Graphs and Results} We have used the following masses in our calculations. The masses of the particles are $m_{\lb}= 5.6195$ GeV, $m_{\tau}= 1.77682$ GeV,  $m_{\mu}=0.10565837 $ GeV, $m_{\lc} = 2.28646$ GeV,
$m_b = 4.18$ GeV,
$m_c = 1.275$ GeV and $V_{cb}=0.0414$ \cite{pdg}.

In the following we present the results for $R_{\lb}$,$\frac{d\Gamma}{dq^2}$ and $B_{\lb}(q^2)$. For the first and third observables we use different models of the form factors
given in Table.\ref{FF}. For the differential distribution
$\frac{d\Gamma}{dq^2}$ we present the average result over the form factors.
 \begin{table}[tbh]
\center
\begin{tabular}{|c|c|c|c|c|c|c|c|c|}\hline
continuum\ model & 1 & 2 & 3 & 4 & Average & Ref.~\cite{Gutsche:2015mxa} & Ref.~\cite{Woloshyn:2014hka} &Ref.~\cite{Detmold:2015} \\
\hline
$R_{\Lambda_{b}}(SM)$ & 0.31  & 0.29 & 0.28 & 0.28 & $0.29\pm .02$ &0.29 &0.31 &$0.34\pm .01$ \\
\hline
\end{tabular}
\caption{Values of $R_{\Lambda_{b}}$ in the SM}
\label{SMpredictions}
\end{table}

We first present our prediction for $R_{\lb}$ in the SM, in Table.~\ref{SMpredictions}, for the various choices of the form factors in Table.\ref{FF}. We also compare  our results with other calculations of this quantity
by other groups using different form factors. We find the average value for $R_{\lb}$ in the SM,  $R_{\lb,SM}= 0.29\pm.02$. This agrees very well with values for this quantity obtained in Ref.~\cite{Gutsche:2015mxa} which uses a covariant confined quark model for the form factors, Ref.~\cite{Woloshyn:2014hka} which uses the form factor model in Ref.~\cite{Pervin:2005ve}, and Ref.~\cite{Detmold:2015} which uses the lattice QCD. This confirms our earlier assertion that the ratio $R_{\lb}$ is largely free from form factor uncertainties
making it an excellent probe to find new physics.

\begin{figure}
\begin{center}
\includegraphics[width=6.5cm, height=5.5cm]{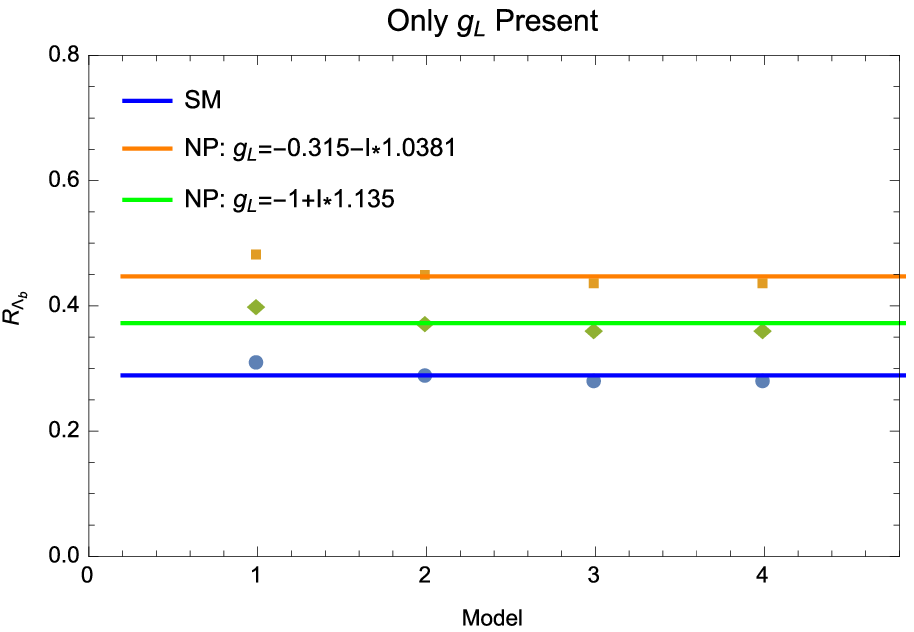}
\includegraphics[width=6.5cm, height=5.5cm]{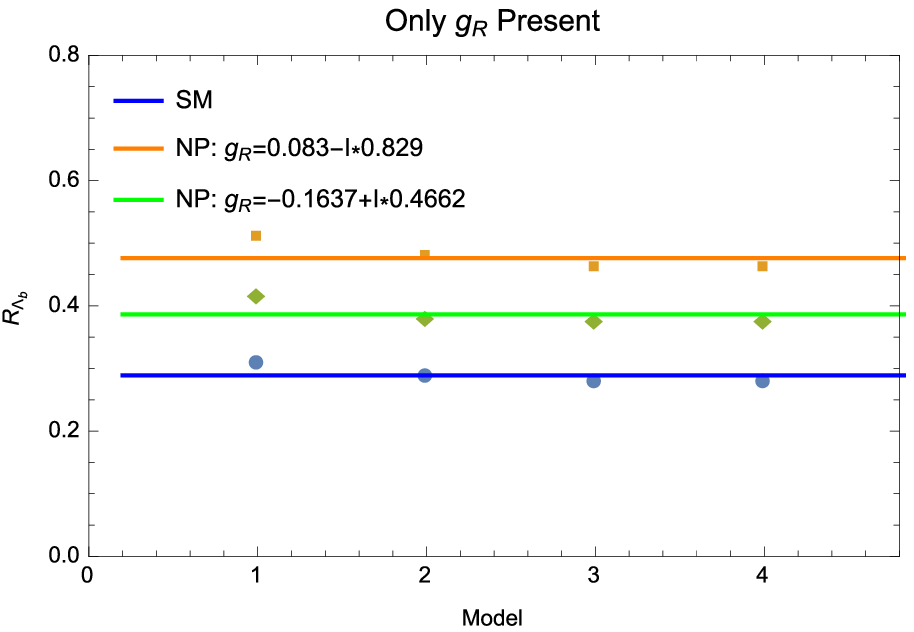}
\includegraphics[width=6.5cm, height=5.5cm]{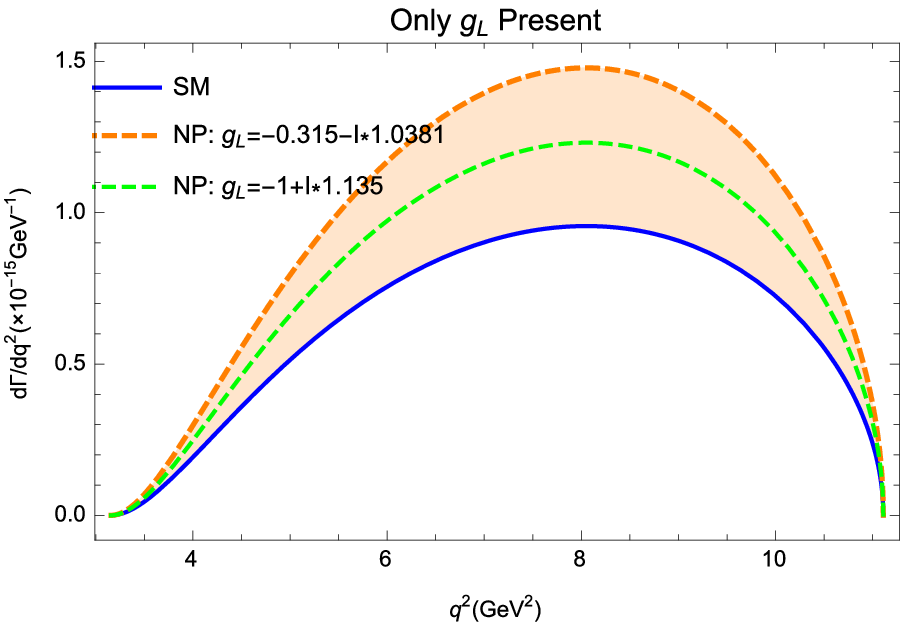}
\includegraphics[width=6.5cm, height=5.5cm]{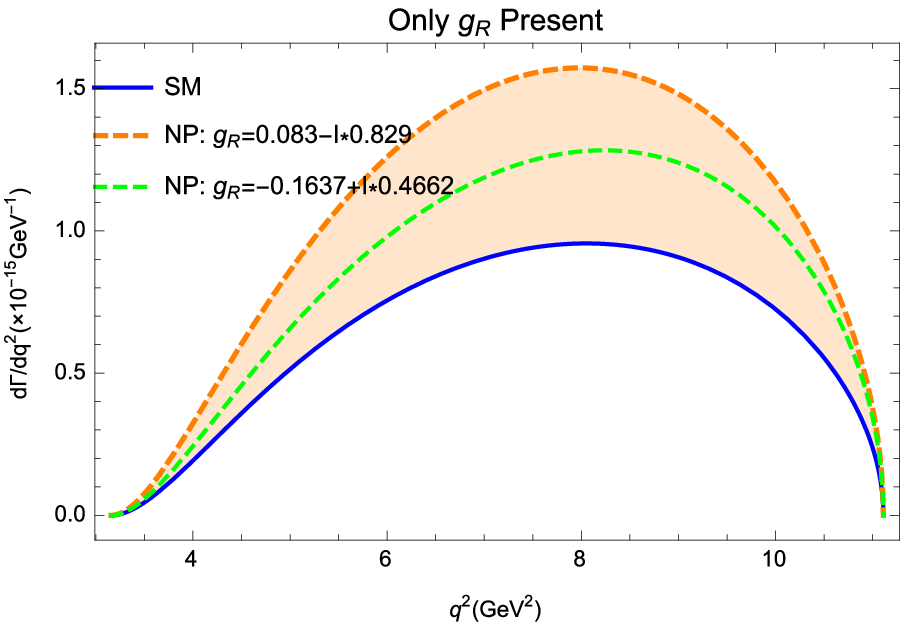}
\includegraphics[width=6.5cm, height=5.5cm]{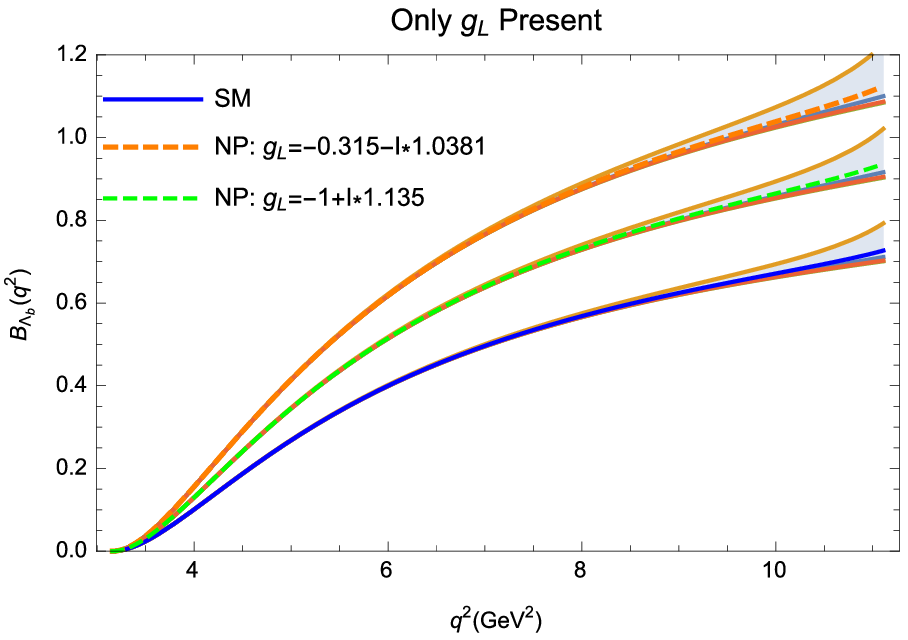}
\includegraphics[width=6.5cm, height=5.5cm]{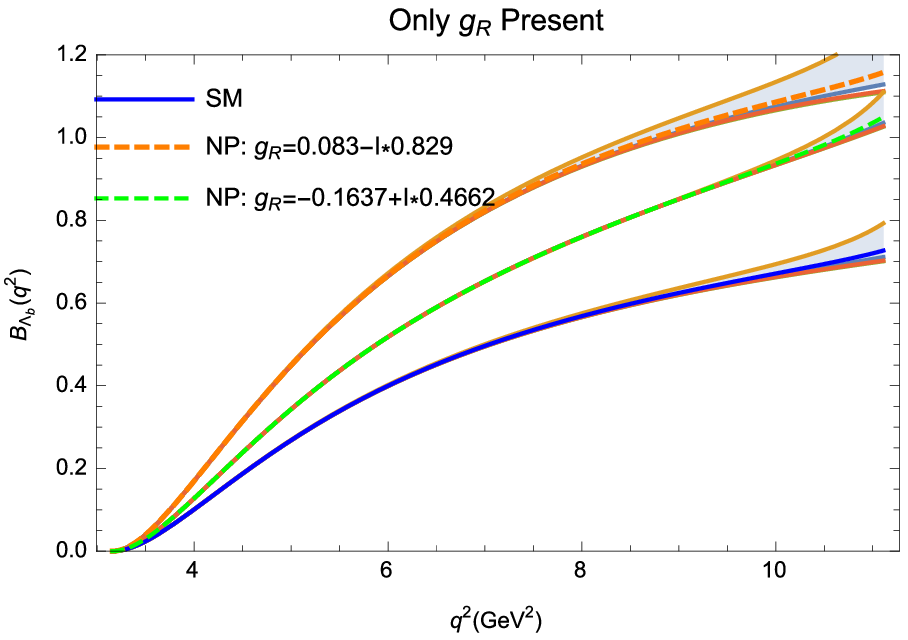}
\end{center}
\caption{The graphs on the left-side (right-side) show the compared results between the standard model and new physics with only $g_L$ ($g_R$) present.  The top and bottom row of graphs depict
$R_{\lb}=\frac{BR[{\Lambda_{b}\to\Lambda_{c}\tau\bar{\nu}_{\tau}}]}{BR[{\Lambda_{b}\to\Lambda_{c}\ell\bar{\nu}_{\ell}}]}$ 
and the
ratio of differential distributions $B_{\lb}(q^2)=\frac{\frac{d\Gamma}{dq^{2}}(\Lambda_{b}\to\Lambda_{c}\tau\bar{\nu}_{\tau})}{\frac{d\Gamma}{dq^{2}}(\Lambda_{b}\to\Lambda_{c}\ell\bar{\nu}_{\ell})}$ as a function of $q^{2}$,
respectively for the various form factors in Table.~\ref{FF}. The middle graphs depict 
the average differential decay rate with respect to $q^2$ for  the process $\Lambda_{b}\to\Lambda_{c}\tau\bar{\nu}_{\tau}$. Some representative values of the couplings have been chosen. 
}
\label{gL}
\end{figure}

We now discuss our results. From the structure of Eq.~\ref{dq1} we can make some general observations.
We start with the case where  only $g_{L}$ is present. In this case the NP has the same structure as the SM and the SM amplitude gets modified by the factor $(1+g_L)$ \cite{Bhattacharya:2014wla}.
Hence, if only $g_L$ is present then
\bea
\label{gLratio}
R_{\lb} & = & R_{\lb}^{SM}|1+g_L|^2. \
\eea
Therefore in this case $R_{\lb} \ge  R_{\lb}^{SM}$ and we find the  range of
$R_{\lb}$ to be $0.44- 0.31$.
 The shape of the differential distribution $\frac{d\Gamma}{dq^2}$ is the same as the SM.
In the left-side figures of Fig.~\ref{gL} we show the plots for $R_{\lb}$,$\frac{d\Gamma}{dq^2}$ and $B_{\lb}(q^2)$ when only $g_L$ is present. We then consider the case where only $g_{R}$ is present.
If only $g_R$ is present then from Eq.~\ref{helicity1},
\bea
\label{gRratio}
H_{\lambda_{\Lambda_{c}},\lambda_{w}}^V & = & (1+g_R) \left[H_{\lambda_{\Lambda_{c}},\lambda_{w}}^V\right]_{SM},\nonumber\\
H_{\lambda_{\Lambda_{c}},\lambda_{w}}^A & = & (1-g_R) \left[H_{\lambda_{\Lambda_{c}},\lambda_{w}}^A\right]_{SM}.\
\eea
In this case no clear relation between $R_{\lb}$ and  $R_{\lb}^{SM}$ can be obtained.
However, for the allowed $g_R$ couplings 
 we find $R_{\lb}$ is greater than the SM value and is in the range $0.47 - 0.30$ .
The shape of the differential distribution $\frac{d\Gamma}{dq^2}$ is the same as the SM.
In the right-side figures of Fig.~\ref{gL} we show the plots for $R_{\lb}$,$\frac{d\Gamma}{dq^2}$ and $B_{\lb}(q^2)$ when
only $g_R$ is present. 

We now move to the case when  only $g_{S,P}$ are present. Using  Eq.~\ref{dq1} and Eq.~\ref{SP} we can write,
\bea
\label{SPratio}
R_{\lb} & = & R_{\lb}^{SM} + |g_P|^2 A_P + 2 Re(g_P) B_P, \nonumber\\
R_{\lb} & = & R_{\lb}^{SM} + |g_S|^2 A_S + 2 Re(g_S) B_S. \
\eea
The quantities $A_{S,P}$ and $B_{S,P}$ depend on masses and form factors and they are positive. Hence for $ Re(g_P) \ge 0$ or  $ Re(g_S) \ge 0$, $R_{\lb}$ is always greater than or equal to  $R_{\lb}^{SM}$. But, for
$ Re(g_P) < 0$ or  $ Re(g_S) < 0$, $R_{\lb}$ can be less than the SM value. However,
given the constraints on $g_{S,P}$ we can make $R_{\lb}$ only slightly less than the SM value. We find $R_{\lb}$ is in the range $0.36- 0.28$ when only $g_S$ is present and
in the range $0.42- 0.30$ when only $g_P$ is present.

\begin{figure}
\begin{center}
\includegraphics[width=6.5cm, height=5.5cm]{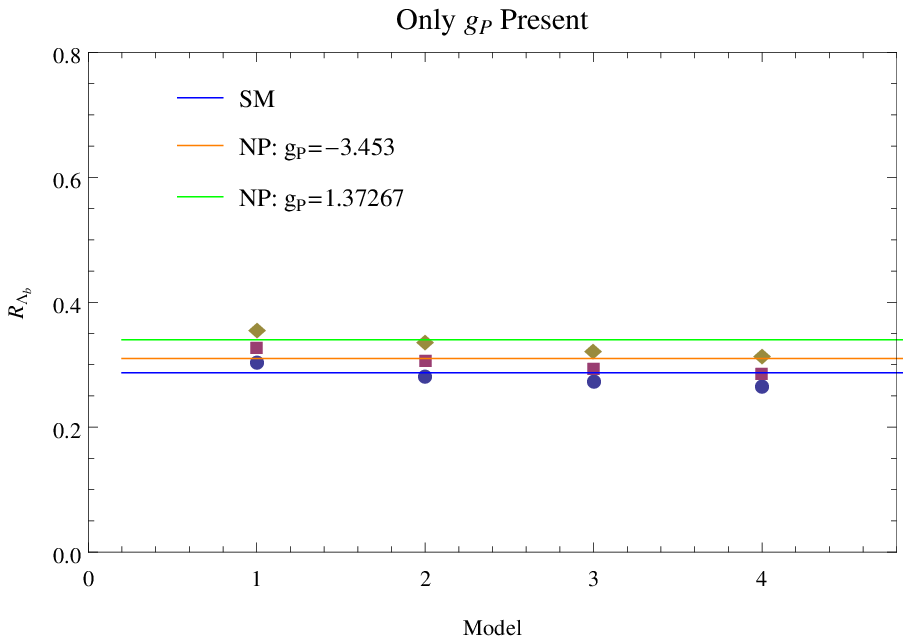}
\includegraphics[width=6.5cm, height=5.5cm]{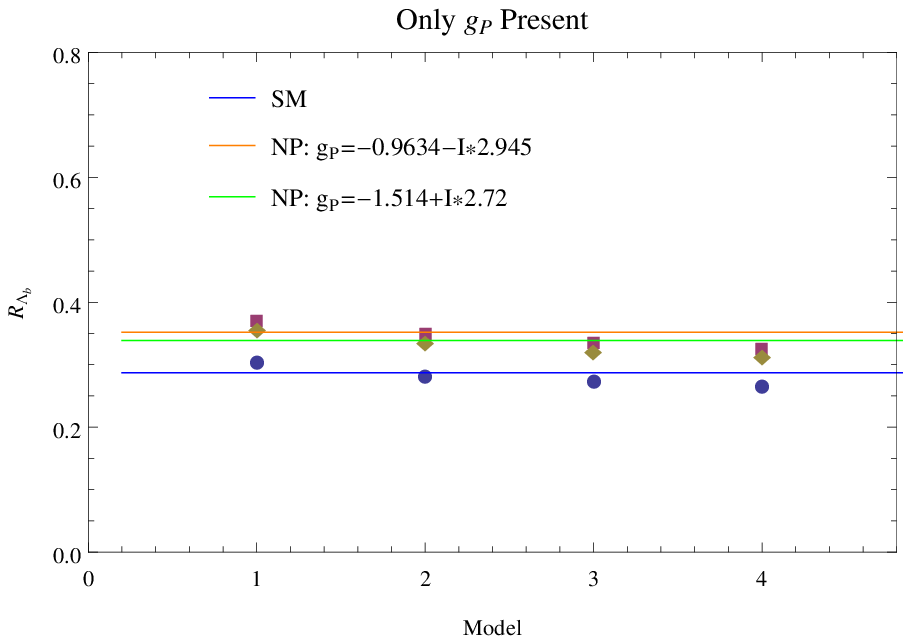}
\includegraphics[width=6.5cm, height=5.5cm]{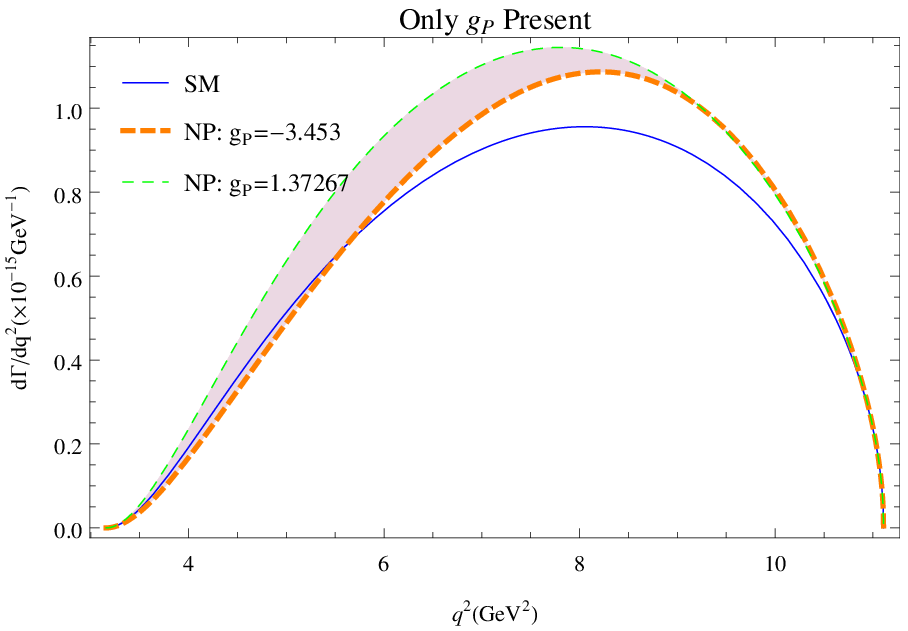}
\includegraphics[width=6.5cm, height=5.5cm]{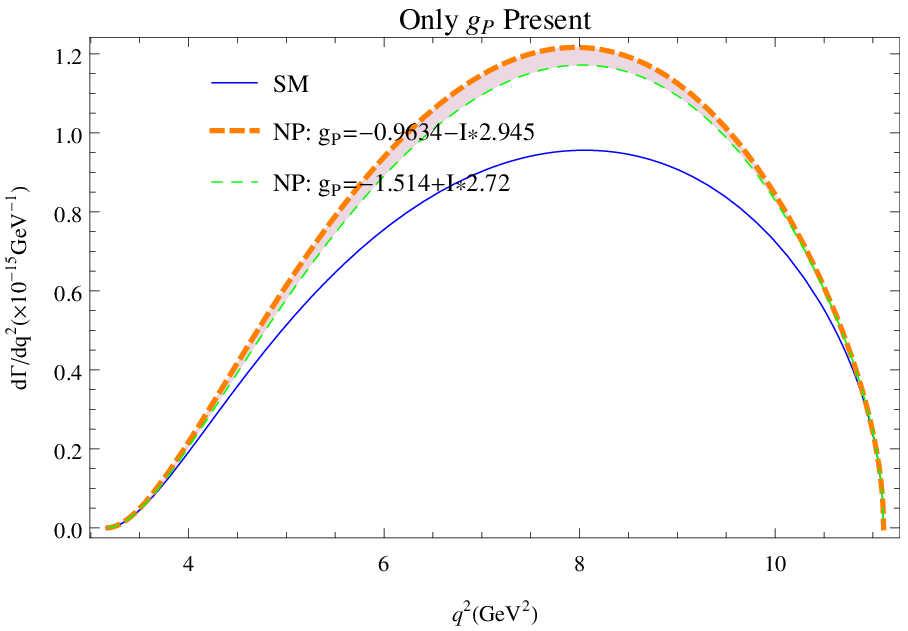}
\includegraphics[width=6.5cm, height=5.5cm]{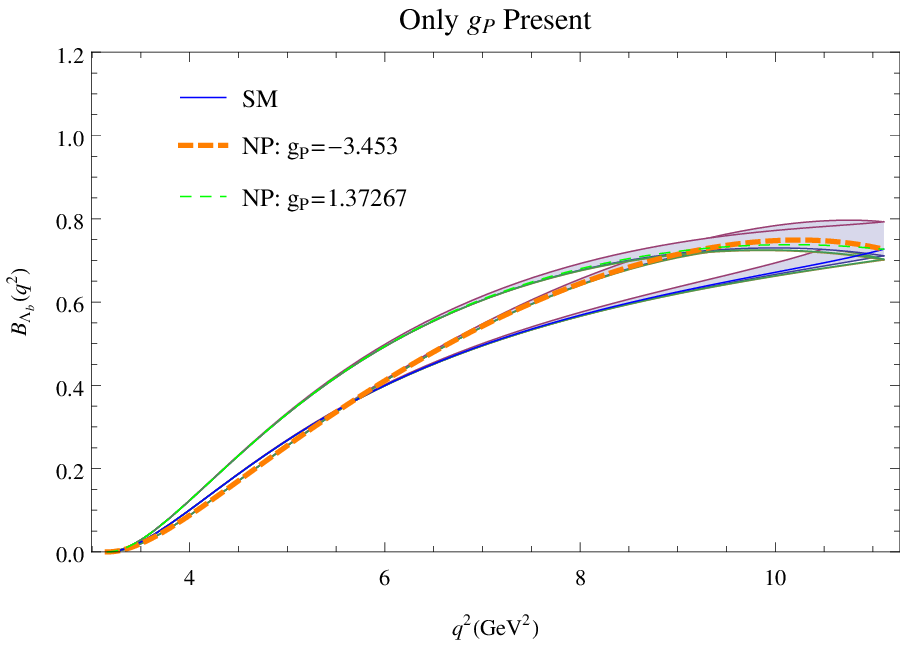}
\includegraphics[width=6.5cm, height=5.5cm]{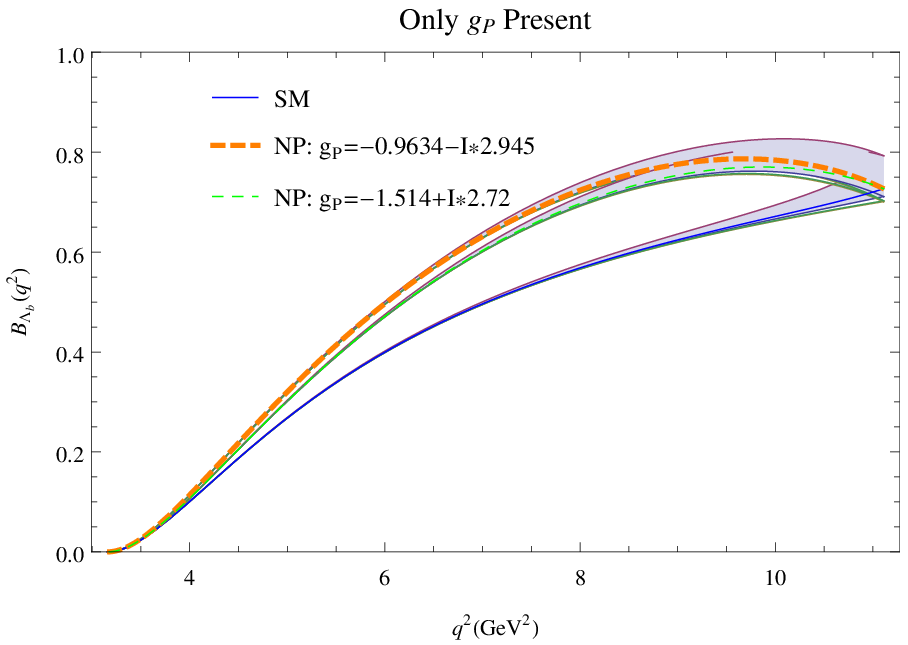}
\end{center}
\caption{The figures show the compared results between the standard model and new physics with only $g_{P}$ present.
The top and bottom row of graphs depict
$R_{\lb}=\frac{BR[{\Lambda_{b}\to\Lambda_{c}\tau\bar{\nu}_{\tau}}]}{BR[{\Lambda_{b}\to\Lambda_{c}\ell\bar{\nu}_{\ell}}]}$ 
and the
ratio of differential distributions $B_{\lb}(q^2)=\frac{\frac{d\Gamma}{dq^{2}}(\Lambda_{b}\to\Lambda_{c}\tau\bar{\nu}_{\tau})}{\frac{d\Gamma}{dq^{2}}(\Lambda_{b}\to\Lambda_{c}\ell\bar{\nu}_{\ell})}$ as a function of $q^{2}$,
respectively for the various form factors in Table.~\ref{FF}. The middle graphs depict 
the average differential decay rate with respect to $q^2$ for  the process $\Lambda_{b}\to\Lambda_{c}\tau\bar{\nu}_{\tau}$. Some representative values of the couplings have been chosen. 
}
\label{gP1}                                                                                                                                                                                                                                                                                                                                                                                                                                                                                                                                                                                                                                 
\end{figure}

\begin{figure}
\begin{center}
\includegraphics[width=6.5cm, height=5.5cm]{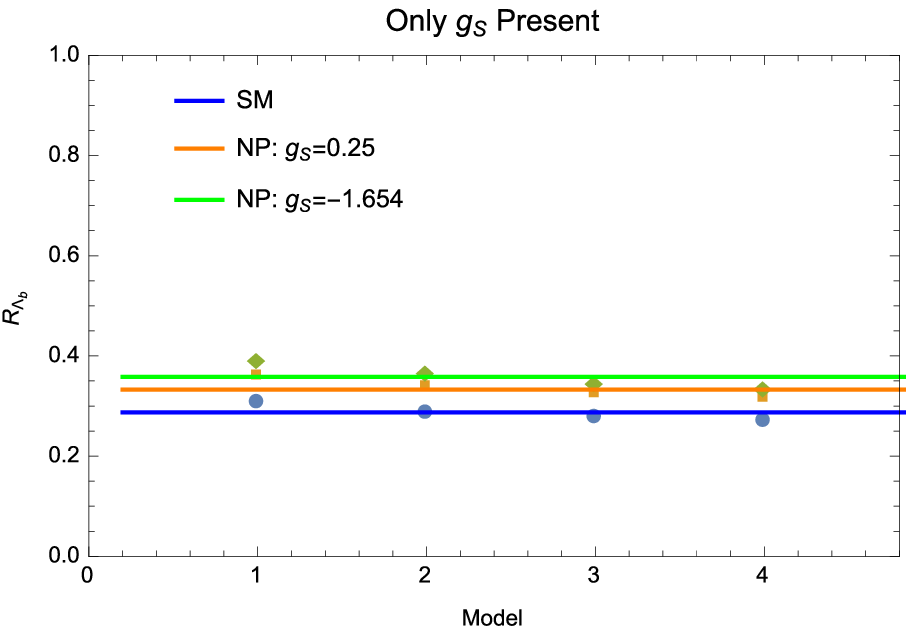}
\includegraphics[width=6.5cm, height=5.5cm]{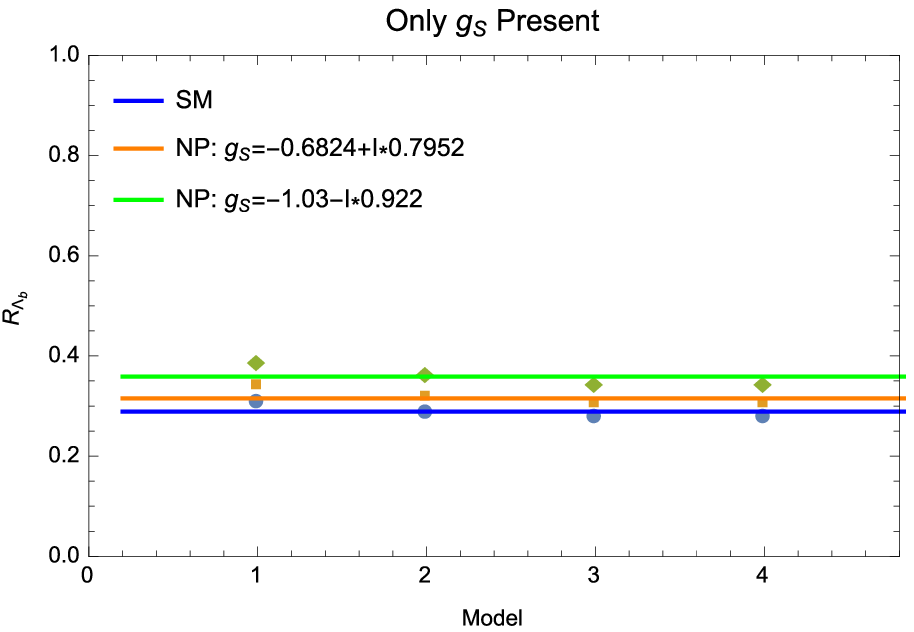}
\includegraphics[width=6.5cm, height=5.5cm]{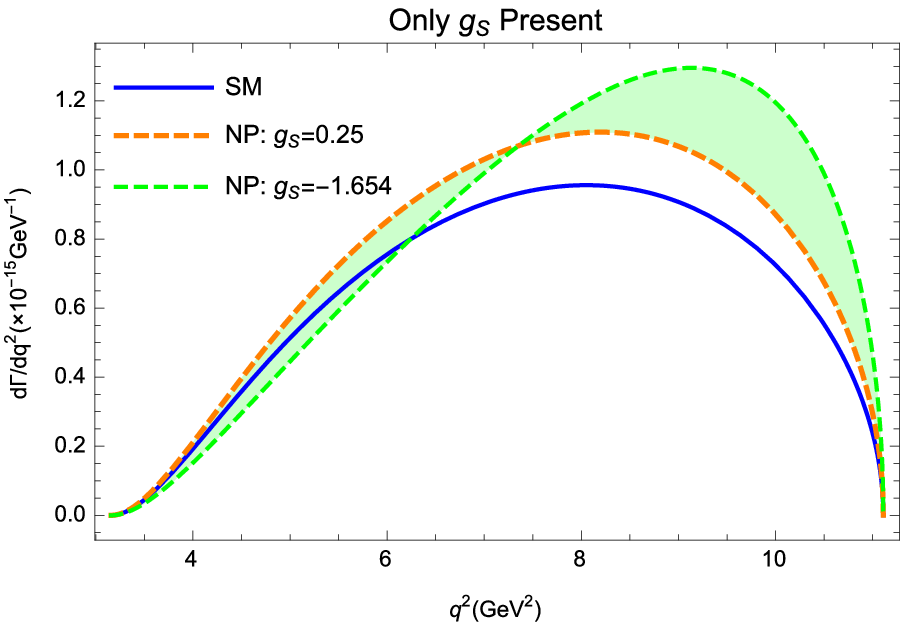}
\includegraphics[width=6.5cm, height=5.5cm]{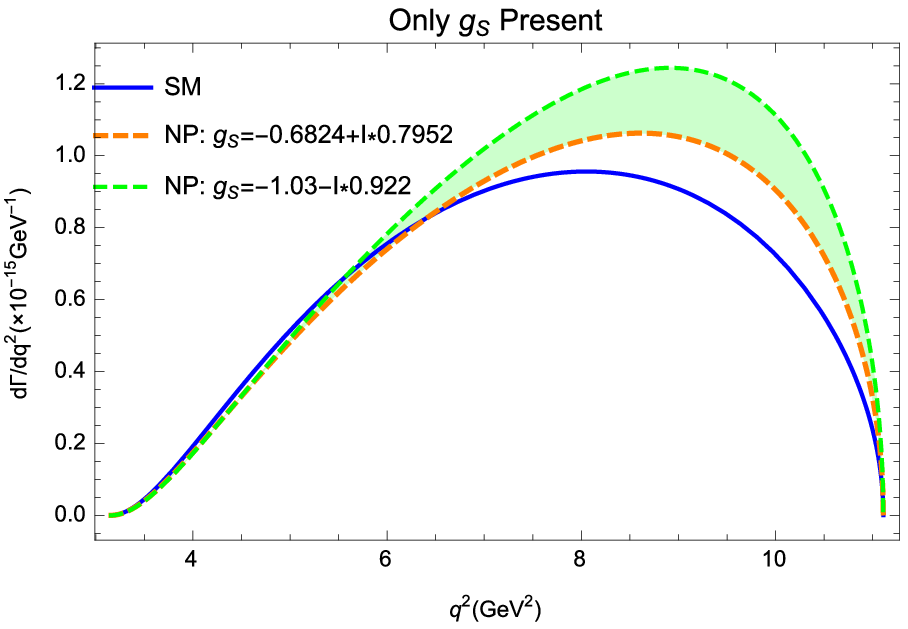}
\includegraphics[width=6.5cm, height=5.5cm]{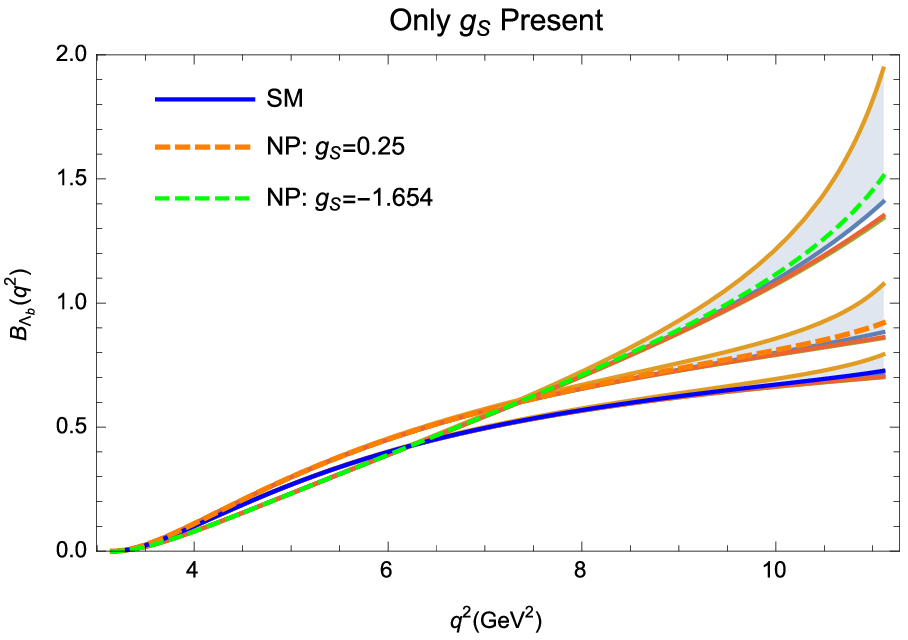}
\includegraphics[width=6.5cm, height=5.5cm]{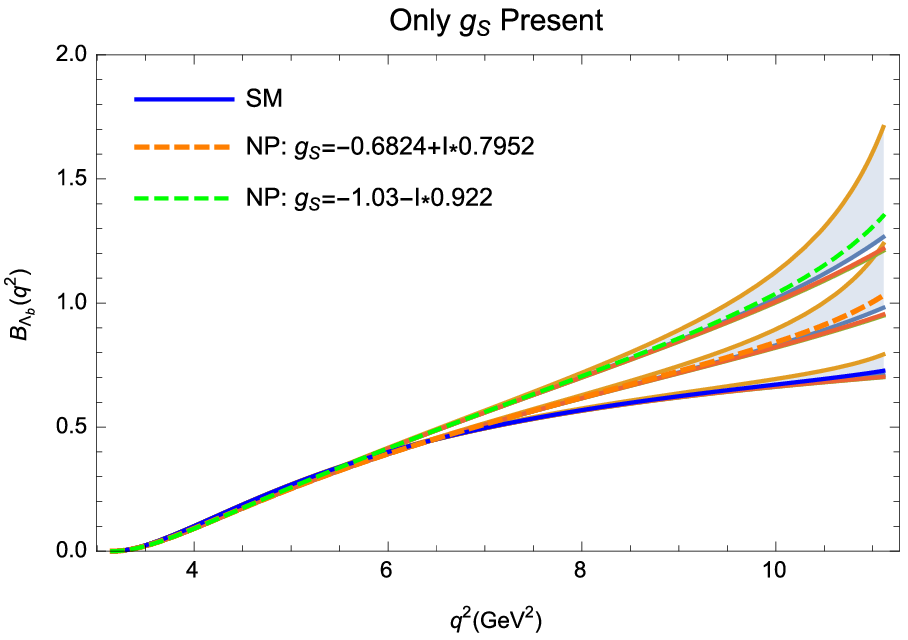}
\end{center}
\caption{
The figures show the compared results between the standard model and new physics with only $g_{S}$ present.
The top and bottom row of graphs depict
$R_{\lb}=\frac{BR[{\Lambda_{b}\to\Lambda_{c}\tau\bar{\nu}_{\tau}}]}{BR[{\Lambda_{b}\to\Lambda_{c}\ell\bar{\nu}_{\ell}}]}$ 
and the
ratio of differential distributions $ B_{\lb}(q^2)=\frac{\frac{d\Gamma}{dq^{2}}(\Lambda_{b}\to\Lambda_{c}\tau\bar{\nu}_{\tau})}{\frac{d\Gamma}{dq^{2}}(\Lambda_{b}\to\Lambda_{c}\ell\bar{\nu}_{\ell})}$ as a function of $q^{2}$,
respectively for the various form factors in Table.~\ref{FF}. The middle graphs depict 
the average differential decay rate with respect to $q^2$ for  the process $\Lambda_{b}\to\Lambda_{c}\tau\bar{\nu}_{\tau}$. Some representative values of the couplings have been chosen. 
}
\label{gS1}
\end{figure}

In  Fig.~\ref{gP1} we show the plots for $R_{\lb}$,$\frac{d\Gamma}{dq^2}$ and $B_{\lb}(q^2)$ when
only $g_P$ is present. The shape of the differential distribution $\frac{d\Gamma}{dq^2}$ can be different from the SM.
In  Fig.~\ref{gS1} we show the plots for $R_{\lb}$,$\frac{d\Gamma}{dq^2}$ and $B_{\lb}(q^2)$ when
only $g_S$ is present. In this case also the shape of the differential distribution $\frac{d\Gamma}{dq^2}$ can be different from the SM.  

\begin{table}[tbh]
\center
\begin{tabular}{|c|c|c|}\hline
NP & $R_{\lb, min} $ & $R_{\lb, max} $ \\
\hline
Only $g_L$ & $0.31 $, $g_L=-0.502 +0.909 \ i $ & $0.44$, $g_L=-0.315 -1.0381 \ i $\\
 Only $g_R$& $0.30 $, $g_R=-0.035 -0.104 \ i $ &   $0.47$, $g_R=0.0827 +0.829 \ i$\\
Only $g_S$ & $0.28 $, $g_S=-0.0227$ & $0.36$, $g_S=-1.66$ \\
Only $g_P$ & $0.30 $, $g_P=0.539$ & $0.42$, $g_P=-5.385$ \\
\hline
\end{tabular}
\caption{Minimum and Maximum values for the averaged $R_{\lb}$.}
\label{Rminmax}
\end{table}

In Table.~\ref{Rminmax} we show the minimum and maximum values for the averaged $R_{\lb}$ with the corresponding NP couplings.

\section{Conclusion}
In this paper we calculated the SM and NP predictions for the decay $\lbt$. Motivation to study this decay comes from the recent hints of
 lepton flavor non-universality observed by the BaBar Collaboration  in $R(D^{(*)}) \equiv
  \frac{{\cal B}({\bar B} \to D^{(*)+} \tau^- {\bar\nu}_\tau)}{ {\cal
    B}({\bar B} \to D^{(*)+} \ell^- {\bar\nu}_\ell)}$ ($\ell =
  e,\mu$).
We used a general parametrization of the NP operators and fixed the new physics couplings from the experimental measurements of $R(D)$ and $R(D^*)$. We then made predictions for $R_{\lb}$ ( Eq.\ref{ratio1}), $\frac{d\Gamma}{dq^2}$, and $B_{\lb}(q^2)$ ( Eq.\ref{ratio2}) for the various NP couplings taken one at a time. We found the interesting result that $g_{L,R}$ couplings gave predictions larger than the SM values for all the three observables. We found the $g_P$ couplings to produce larger effects
than the $g_S$ couplings.
We also provided the general formula for the various angular distributions in the presence of  NP operators.

\bigskip
\noindent
{\bf Acknowledgments}:
This work was financially supported in part by the National Science Foundation 
under Grant No.NSF PHY-1414345. We thank  J$\ddot{\text{u}}$rgen K$\ddot{\text{o}}$rner, Preet Sharma, and Sheldon Stone for useful discussions.

\end{document}